\documentclass[aps,prl,groupedaddress,preprint]{revtex4-1}

\usepackage[dvips]{graphicx}
\usepackage{bm,amsmath,amssymb}

\newcommand{\grad}{{\bm \nabla}}
\newcommand{\rot}{{\bm \nabla} \times}
\renewcommand{\div}{{\bm \nabla} \cdot}
\newcommand{\del}{\partial}
\newcommand{\delt}{\partial_t}
\newcommand{\Delv}{{\bm{\mathcal D}}}
\newcommand{\Delt}{{\mathcal D}_t}

\newcommand{\rv}{{\bm r}}
\newcommand{\kv}{{\bm k}}
\newcommand{\qv}{{\bm q}}
\newcommand{\w}{\omega}
\newcommand{\W}{\Omega}

\newcommand{\A}{A^{\rm em}}
\newcommand{\Av}{{\bm A}^{\rm em}}
\newcommand{\At}{\phi}
\newcommand{\As}{{\mathcal A}}
\newcommand{\Asv}{{\bm{\mathcal A}}}
\newcommand{\Ast}{{\mathcal A}_t}

\newcommand{\Es}{{\mathcal E}}
\newcommand{\Esv}{{\bm{\mathcal E}}}
\newcommand{\Bs}{{\mathcal B}}
\newcommand{\Bsv}{{\bm{\mathcal B}}}
\newcommand{\jv}{{\bm j}}
\newcommand{\sv}{{\bm s}}

\newcommand{\Iv}{{\bm I}}

\newcommand{\Isv}{{\bm{\mathcal I}}}

\newcommand{\Ev}{{\bm E}}
\newcommand{\Bv}{{\bm B}}

\newcommand{\Btv}{\tilde{\bm B}}

\newcommand{\pauli}{\hat{\sigma}}

\newcommand{\vi}{v_{\rm i}}
\renewcommand{\ni}{n_{\rm i}}
\newcommand{\ui}{u_{\rm i}}
\newcommand{\Ef}{\varepsilon_{\rm F}}
\newcommand{\kf}{k_{\rm F}}

\newcommand{\mub}{\mu_{\rm B}}
\newcommand{\cond}{\sigma_{\rm c}}
\newcommand{\chiL}{\chi_{\rm L}}

\newcommand{\tr}{{\rm tr}}
\newcommand{\G}{\hat{G}}

\renewcommand{\Im}{{\rm Im}}
\newcommand{\ga}{g^{\rm a}}
\newcommand{\gr}{g^{\rm r}}

\newcommand{\Ek}{\varepsilon_{\bm k}}

\renewcommand{\a}{\alpha}
\newcommand{\av}{{\bm \alpha}}

\newcommand{\yv}{{\bm y}}
\newcommand{\zv}{{\bm z}}

\begin{document}

%%%%%%%%%%
\title{Theory of Electrical Spin Manipulation in Spin-Orbit Coupling Systems}
%%%%%

\author{Akihito Takeuchi$^1$}
\email[]{takeuchi@appi.t.u-tokyo.ac.jp}

\author{Naoto Nagaosa$^{1,2}$}

%\homepage[]{Your web page}
%\thanks{}
%\altaffiliation{}

\affiliation{
$^1$Department of Applied Physics, University of Tokyo, Hongo, Tokyo 113-8656, Japan
\\
$^2$RIKEN Center for Emergent Matter Science, Wako, Saitama 351-0198, Japan
}

%%%%%

\date{\today}

%%%%%%%%%%

\begin{abstract}
By associating a spin-orbit interaction with a non-Abelian gauge potential,
we theoretically present a spin polarization in a quite general form using an effective Yang-Mills field and a usual electromagnetic field.
In this gauge invariant result, we focus on a purely electrically-induced spin contribution.
We find that both the inverse spin galvanic effect and the spin Hall effect arise from the same origin, i.e., the SU(2)$\times$U(1) Hall effect.
We also discover that a large effective magnetic field of the order of $1$T is induced in the Rashba system.
\end{abstract}

%%%%%%%%%%

%\pacs{}

%\keywords{}

%%%%%%%%%%

\maketitle

Generation of spin by applying an electric current in the presence of spin-orbit interaction has been investigated with much theoretical and experimental attention in spintronics~\cite{Wolf01, Zutic04}.
One of the most successful phenomena of electronic spin-and-charge coupled transport is the spin Hall effect~\cite{Hirsch99, Murakami03, Sinova04, Sinitsyn04, Kato04}.
In a system with spin-orbit interaction, a spin current appears in the transverse direction to an applied electric current.
As a result, electronic spin accumulates at the edges of the sample.
As a similar effect, in the inverse of the spin galvanic effect~\cite{Edelstein90, Ganichev02, Ganichev04, Ganichev06} a spin polarization is also induced by applying an electric current.
These two phenomena are different in a direction of an emergent spin polarization.
In a case of a two-dimensional electron system without inversion symmetry, the induced spin polarization is out-of-plane in the spin Hall effect, while the in-plane spin arises in the inverse spin galvanic effect.

Although the electronic spin is the well-defined quantity, the theoretical definition of spin current is not uniquely given under the spin-orbit interaction.
In the presence of spin-orbit interaction, electronic spin dynamics always accompanies the relaxation compared with the equation of motion for electric charge.
To resolve this ambiguity in the definition, the non-Abelian gauge theory is one of the possible solutions.
To connect the spin-orbit coupling with the non-Abelian gauge theory in condensed matters has been the well-known idea for many years~\cite{Frohlich93}, and a proper definition of spin current is given on the basis of the SU(2) gauge invariance by treating the spin-orbit interaction as the non-Abelian vector potential~\cite{Leurs08, Tokatly08, Duckheim09}.
In this context, despite the conservation law for spin is still broken, the electronic spin is covariantly conserved,
\begin{equation}
\delt s^a +\div \jv^a
=
-\frac{2 e}{\hbar} \epsilon^{abc} (\Ast^b s^c -\Asv^b \cdot \jv^c),
\label{eq:spin-conti}
\end{equation}
where $s$ is the electronic spin, $j_i^a$ is the spin current flowing in the $i$-direction and spin-polarized in the $a$-direction, and $\As$ represents the non-Abelian spin-orbit gauge potential.
In recent years, several spin-dependent phenomena based on this non-Abelian gauge theory have been actively reported~\cite{Frohlich93, Leurs08, Tokatly08, Duckheim09, Bernevig06, Hatano07, Gorini10, Sugimoto12}.

Following the non-Abelian gauge theory, time and space derivatives of the spin-orbit coupling are corresponding to effective Yang-Mills electric and magnetic fields, respectively, which drive spin current and spin polarization.
In experiment, a spatial and temporal variation of the spin-orbit coupling is feasible.
For example, since the Rashba effect emerges when the gate voltage breaks the structural inversion symmetry in two-dimensional semiconductor heterostructures~\cite{Rashba60, Nitta97}, the alternating gate voltage could change the Rashba coupling, and in the specific sample configuration the spatially-varying Rashba coupling is realized~\cite{Kohda12}.
This space-time dependent spin-orbit coupling is expected to open up the possibilities of electrical spin manipulation.

In this paper, we derive analytically a general expression of spin polarization in terms of an effective non-Abelian SU(2) Yang-Mills field corresponding to the spin-orbit interaction and Zeeman effect, and the usual U(1) Maxwell electromagnetic field.
In particular, we focus on the generation of spins by electric field alone without any magnetic contributions.
A related work has been done by Gorini {\it et al.} who demonstrated theoretically a SU(2)$\times$U(1) covariant Boltzmann equation in a space-time dependent two-dimensional Rashba system, and derived explicitly electronic spin and charge transport~\cite{Gorini10}.
The focus of their work, however, is the spin and charge currents in contrast to the spin polarization in the present paper.

We consider a general disordered electron system coupled to an external electromagnetic field and a spin-orbit interaction in condensed matters whose Hamiltonian is represented by
\begin{align}
H
=
&\frac{1}{2 m} \int{d^d r}
\big| \big[ -i \hbar \grad +e \Av(\rv,t) +e \Asv^a(\rv,t) \pauli^a \big] \Psi(\rv,t) \big|^2
\notag
\\
&-e \int{d^d r}
\Psi^\dagger(\rv,t) \big[ \At(\rv,t) +\Ast^a(\rv,t) \pauli^a \big] \Psi(\rv,t)
\notag
\\
&-\Ef \int{d^d r}
\Psi^\dagger(\rv,t) \Psi(\rv,t)
+H_{\rm i},
\label{eq:Hamiltonian}
\end{align}
where
$\Psi = (\psi_\uparrow, \psi_\downarrow)$ is the annihilation operator of conduction electron,
$d$ denotes the number of dimensions,
$m$ and $-e$ are mass and charge of electron, respectively,
$\hbar$ is the Planck constant,
$\Ef$ is the Fermi energy,
$\pauli^a$ is the vector of the Pauli matrices ($a = x,y,z$ and the caret means a matrix),
and $H_{\rm i}$ denotes the spin-independent random impurity scattering which gives rise to the relaxation time of electron, $\tau$.
Here, the external electric and magnetic fields are defined using the potential $\phi$ and $\Av$ as $\Ev = -\delt \Av -\grad \At$ and $\Bv = \rot \Av$, respectively.
The non-Abelian spin-orbit gauge potential $\As_\mu^a$ ($\mu = t,x,y,z$) is the first-order relativistic correction of electromagnetic field derived from the Dirac equation, and its time and space components are consistent with the Zeeman splitting and the spin-orbit coupling, respectively,
\begin{align}
\Ast^a
&=
-\frac{\hbar}{2 m} B^a,
&
\As_i^a
&=
\frac{\hbar}{4 m c^2} \epsilon_{ija} E_j,
\label{eq:def-gauge}
\end{align}
where $c$ is the speed of light and $\epsilon_{ija}$ is the antisymmetric tensor.
According to the non-Abelian gauge theory, the effective Yang-Mills electric field is defined as
$\Esv^a = -\delt \Asv^a -\grad \Ast^a -(2 e / \hbar) \epsilon^{abc} \Ast^b \Asv^c$,
and its magnetic counterpart is
$\Bsv^a = \rot \Asv^a -(e / \hbar) \epsilon^{abc} \Asv^b \times \Asv^c$.

\begin{figure}
\includegraphics[scale=0.9, clip]{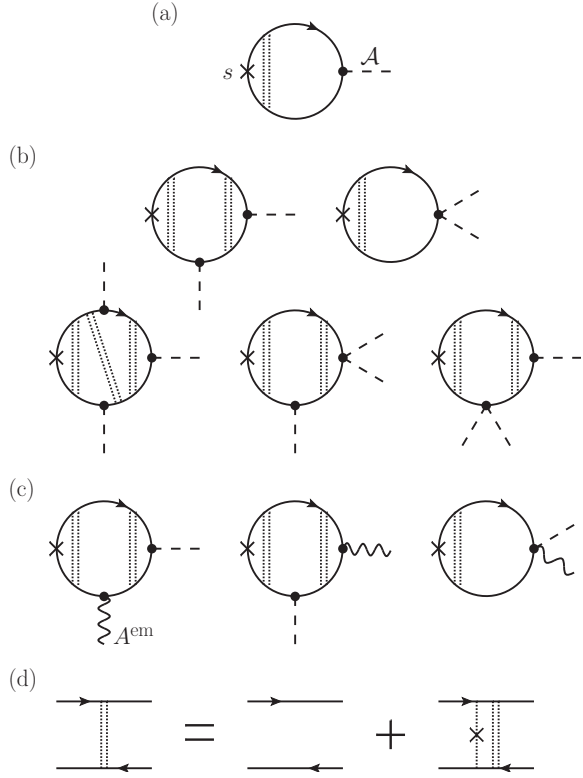}
\caption{Diagrammatic representation of spin density and vertex correction.
The solid line denotes the impurity-averaged Green's function of free electrons, the dashed line represents interaction with the SU(2) spin-orbit gauge potential, $\As$, the wavy line describes the interaction with general U(1) electromagnetic potential, $\A$.
The double dotted line is the diffusion ladder due to the impurity scattering described by the single dotted line.}
\label{fig:spin}
\end{figure}

From the covariant spin conservation law [Eq.~\eqref{eq:spin-conti}], we can introduce a covariantly conserved spin current,
\begin{align}
j_i^a
=
&-\frac{i e \hbar}{2 m} \tr
\big(
\langle\!\langle{\nabla_i \Psi^\dagger \pauli^a \Psi}\rangle\!\rangle
-\langle\!\langle{\Psi^\dagger \pauli^a \nabla_i \Psi}\rangle\!\rangle
\big)
\notag
\\
&-\frac{e^2}{m} \tr
\big(
\A_i \langle\!\langle{\Psi^\dagger \pauli^a \Psi}\rangle\!\rangle
+\As_i^a \langle\!\langle{\Psi^\dagger \Psi}\rangle\!\rangle
\big),
\end{align}
where
$\tr$ denotes trace over spin indices and
$\langle\!\langle{\cdots}\rangle\!\rangle$ is the quantum expectation value.
Using the Keldysh Green's function~\cite{Rammer86, Haug07}, we carry out the analytic calculation of the spin and its current densities induced by the effective Yang-Mills field and electromagnetic field.
The electronic spin density is generally defined as $s^a(\rv,t) = i e \hbar \tr [ \pauli^a \G^<(\rv,t; \rv,t) ]$, where $G^<_{\sigma, \sigma'}(\rv,t; \rv',t') = (i / \hbar) \langle\!\langle{\psi^\dagger_{\sigma'}(\rv',t') \psi_\sigma(\rv,t)}\rangle\!\rangle$ is the lesser component of the Keldysh Green's function.

For calculation, we show Feynman diagrams of the spin density in Fig.~\ref{fig:spin}.
We consider slowly-varying electric and magnetic fields, subject to $k \ell \ll 1$ and $\w \tau \ll 1$ ($k$ and $\w$ are wavenumber and frequency of electromagnetic field, and $\ell$ denotes a mean free path of electrons), and we do the gradient expansion.
The calculation on the basis of the quantum many-body theory (see Supplemental Material in detail) yields the following result up to the second order in $\A$ and $\As$ (including the third-order terms which guarantees the gauge covariance),
\begin{align}
s^a
=
&-\cond \div \langle{\Esv^a}\rangle
+\frac{2 e}{\hbar} \epsilon^{abc} \div \langle{\chiL \Asv^b \times \Bsv^c -D \Asv^b s^c}\rangle
\notag
\\
&-\frac{e \tau}{m}
\div
\langle{
\Bv \times \Isv^a
+\Bsv^a \times \Iv
-S^a \Ev
-\rho \Esv^a
}\rangle
\notag
\\
&+\frac{e \tau}{m} \grad^2
\langle{e^2 \nu \At \Ast^a +\At S^a +\rho \Ast^a}\rangle
\notag
\\
&-\frac{2 e}{\hbar} \epsilon^{abc} \langle{\As_t^b s^c -\Asv^b \cdot \jv^c}\rangle.
\label{eq:spin}
\end{align}
This spin polarization is clearly formed in terms of the effective SU(2) Yang-Mills field due to spin-orbit coupling and the U(1) Maxwell electromagnetic field.
This simple equation is one of the main conclusions in the present paper.
In Eq.~\eqref{eq:spin}, the angle bracket denotes the average of a diffusive electron motion satisfying the relation $(-D \grad^2 +\delt) \langle{F}\rangle = F$ ($F$ is the arbitrary function with respect to space and time),
$\cond = e^2 \nu D$ is the conductivity,
$\nu$ is the density of states per volume involving spin degree of freedom,
$D = 2 \Ef \tau / d m$ is the diffusion constant,
$\chiL  = -\mub^2 \nu / 3$ is the Landau diamagnetic susceptibility,
and $\mub = e \hbar / 2 m$ is the Bohr magneton.
The spin polarization and the spin current driven by the Yang-Mills electric field $\Es$ are defined as 
$S^a = -\cond \langle{\div \Esv^a}\rangle$ and $\Isv^a = \cond \Esv^a -D \grad S^a$, respectively,
and as their U(1) counterparts the electric charge and electric current are given by $\rho = -\cond \langle{\div \Ev}\rangle$ and $\Iv = \cond \Ev -D \grad \rho$.
In a similar manner, the spin current is also obtained,
\begin{align}
\jv^a
=
&\cond \Esv^a
+\chiL \rot \Bsv^a
-\frac{2 e}{\hbar} \epsilon^{abc}
(\chiL \Asv^b \times \Bsv^c -D \Asv^b s^c)
\notag
\\
&+\frac{e \tau}{m}
(\Bv \times \Isv^a +\Bsv^a \times \Iv -S^a \Ev -\rho \Esv^a)
\notag
\\
&-\frac{e \tau}{m} \grad
(e^2 \nu \At \Ast^a +\At S^a +\rho \Ast^a)
-D \grad s^a.
\label{eq:current}
\end{align}
This result of spin current is consistent with the previous theoretical works~\cite{Tokatly08, Gorini10}.
The spin polarization and the spin current given by Eqs.~\eqref{eq:spin} and \eqref{eq:current}, of course, satisfy the spin continuity equation [Eq.~\eqref{eq:spin-conti}].
We note that the electronic spin is exactly conserved at the linear-order SU(2) gauge potential.
In other words, spin is conserved in a weak spin-orbit coupling system.

Equation~\eqref{eq:spin} is a really compact equation of spin polarization; however, the effective Yang-Mills field, $\Es$ and $\Bs$, is inappropriate to explain the real spin-related phenomena.
Here we replace the non-Abelian spin-orbit gauge potential by the real electromagnetic field using Eq.~\eqref{eq:def-gauge}, and therefore the spin polarization in Eq.~\eqref{eq:spin} is rewritten as
\begin{align}
\sv
=
&e \nu \mub
\bigg[
\Bv
-\langle{\delt \Btv}\rangle
-\frac{e}{m} \big\langle{\Bv \times \Btv}\big\rangle
-\frac{e D}{4 m c^4} \langle{\Ev \times \delt \Ev}\rangle
\notag
\\
&-\frac{e}{m} \big\langle{\Btv \times \langle{\delt \Btv}\rangle}\big\rangle
-\frac{e D^2}{2 m c^2} \big\langle{(\Ev \times \grad) \times \langle{\grad^2 \Btv}\rangle}\big\rangle
\bigg]
\notag
\\
&+\frac{\mub \tau}{2 m c^2}
\bigg\{
\cond \big\langle{(\rot \Bv -\Bv \times \grad) \times \delt \Ev}\big\rangle
-\nabla_i \langle{\nabla_i \Ev \times \Iv}\rangle
\notag
\\
&-\delt \big\langle{\big( \rot \Ev -\Ev \times \grad \big) \rho}\big\rangle
-\nabla_i \big\langle{\big( \rot \Ev -\Ev \times \grad \big) I_i}\big\rangle
\notag
\\
&+\frac{2 c^2}{D} \nabla_i \langle{I_i \Bv}\rangle
-2 e^2 \nu c^2 \nabla_i
\Big[
\big\langle{\big( \Ev +D \rot \Bv \big)_i \langle{\delt \Btv}\rangle}\big\rangle
\notag
\\
&+\nabla_i \big\langle{\At \langle{\delt \Btv}\rangle}\big\rangle
\Big]
\bigg\},
\label{eq:spin_elemag}
\end{align}
where $\Btv \equiv \Bv -(D / 2 c^2) \rot \Ev$ is the magnetic field including the spin-orbit correction.
This is a general form of spin polarization induced by electric and magnetic fields in the presence of spin-orbit interaction.
The first term is due to the Zeeman field.
The next five terms represent the generalized spin pumping effect.
In the usual spin pumping effect~\cite{Silsbee79, Tserkovnyak02, Mizukami02}, dynamic magnetization or magnetic field induces a flow of spin angular momentum and it is a purely magnetic effect.
We here revealed that the electrical counterpart of the spin pumping effect also occurs thanks to the spin-orbit coupling which converts orbital energy into spin.
The origin of the spin pumping effect is the spin torque.
The usual spin pumping effect comes from a precession of electronic spin around a magnetic field, $\epsilon^{abc} \Ast^b s^c$.
In contrast, the origin of the electrical spin pumping effect is a torque owing to the spin-orbit coupling proportional to $\epsilon^{abc} \Asv^b \cdot \jv^c$.

Since electrical control of spin is quite important for developing spintronics, we focus on the purely electrical manipulation of electronic spin.
We pick out the electrically induced spin polarization in Eq.~\eqref{eq:spin_elemag} and it reads
\begin{align}
\sv
=
&-\frac{\cond \mub}{4 m c^4}
\langle{\Ev \times \delt \Ev}\rangle
+\frac{\mub \tau}{2 m c^2}
\big(
\big\langle{\delt \Ev \times \grad \rho}\big\rangle
\notag
\\
&+\langle{\nabla_i \Ev \times \grad I_i}\rangle -\nabla_i \langle{\nabla_i \Ev \times \Iv}\rangle
\big).
\label{eq:spin_ele}
\end{align}
Here we ignored the rotation of electric field because the alternating magnetic field must be necessarily applied to generate a rotational electric field based on the Faraday's law.
Equation~\eqref{eq:spin_ele} expresses a generation of spin by space-time dependent electric field and this is the second main result in this paper.
The first term represents the electrical spin pumping effect and the other contributions originate from the Lorentz force of the effective Yang-Mills electric and magnetic fields, $s^a = (e \tau / m) \div \langle{\Iv \times \Bsv^a +\rho \Esv^a}\rangle$.
Assuming a case of uniform or stationary electric field, Eq.~\eqref{eq:spin_ele} is useful to demonstrate various spintronic phenomena shown in the following.

Let us first consider the spatially uniform case.
In this situation, the induced spin is due to the electrical spin pumping effect shown in the first term of Eq.~\eqref{eq:spin_ele}.
Since this is due to the second order of spin-orbit coupling, it is usually very weak and hard to detect this phenomenon.
As a material effect, the Rashba spin-orbit interaction is a candidate for realizing experimental observation of the electrical spin pumping effect because the really huge Rashba coupling emerges at a boundary or surface of metals~\cite{LaShell96, Ast07, Nakagawa07} and bulk Rashba semiconductors~\cite{Ishizaka11, Bahramy11}.
The Rashba effect is given by ($\a$ is the Rashba coupling)
\begin{equation}
\As_i^a
=
\frac{m}{e \hbar} \epsilon_{ija} \a_j.
\end{equation}
If we consider two kinds of Rashba effect: the bulk Rashba effect along the $z$-axis, $\av_{\rm b} \parallel \zv$, and the time-evolving Rashba effect by the gate voltage in the $y$-direction, $\av_{\rm t} \parallel \yv$, the induced spin polarization is lying in the $x$-direction,
\begin{equation}
s^x(t)
=
\frac{m \cond}{\hbar^2 \mub}
\a_{\rm b} \a_{\rm t}(t).
\end{equation}
We set a weak Rashba coupling in a metallic sample, where $\a_{\rm b} \kf / \Ef \sim 0.1$, $\a_{\rm t} \kf / \Ef \sim 0.1$ ($\Ef \sim 1${eV} and $\kf \sim 1${\AA}$^{-1}$ being the Fermi wavenumber), and $\hbar / \Ef \tau \sim 0.001$.
Then the emergent spin polarization is roughly estimated at $|(\mu_0 \gamma \hbar / 2 e) s^x| \sim 1${T} as an effective magnetic field, where $\mu_0$ is the magnetic permeability in vacuum and $\gamma$ is the gyromagnetic ratio.
This effective magnetic field is quite large enough to control magnetization.
As a related study, it was theoretically proposed that a spin is driven by applying two orthogonal gate electric fields on the different sections of a one-dimensional wire~\cite{Avishai10}.

Next, we investigates the case of steady state in Eq.~\eqref{eq:spin_ele}.
Here we consider two-dimensional Rashba and Dresselhaus systems given by
$\As_i^z = \As_z^i = 0$ ($i = x, y, z$) and
\begin{equation}
\begin{pmatrix}
\As_x^x & \As_x^y
\\
\As_y^x & \As_y^y
\end{pmatrix}
=
\frac{m}{e \hbar}
\begin{pmatrix}
\beta & -\alpha
\\
\alpha & -\beta
\end{pmatrix},
\end{equation}
where $\beta$ represents the Dresselhaus coupling~\cite{Dresselhaus55}.
To assume the injection of electric current along the $y$-direction in the present system, the in-plane spin polarization emerges
\begin{equation}
\begin{split}
s^x(\qv+\qv')
&=
\frac{m}{\hbar \Ef}
\frac{q_x +q_x'}{|\qv+\qv'|^2}
\big[ q_y \beta(\qv) -q_x \alpha(\qv) \big] I(\qv'),
\\
s^y(\qv+\qv')
&=
\frac{m}{\hbar \Ef}
\frac{q_x +q_x'}{|\qv+\qv'|^2}
\big[ q_x \beta(\qv)  -q_y \alpha(\qv) \big] I(\qv').
\end{split}
\end{equation}
The effect is very sensitive to the spatial dependence of the Rashba and Dresselhaus spin-orbit couplings and the external electric current.
The result becomes changed by whether we take first the limit to the constant spin-orbit coupling, $q \rightarrow 0$, or the uniform electric current, $q' \rightarrow 0$.
(i) When we inject the uniform electric current in the spatially-varying Rashba and Dresselhaus systems, the spin polarization turns to ($\grad \a \times \zv = \grad \beta \times \zv = 0$)
\begin{align}
s^x(\rv)
&=
-\frac{m I}{\hbar \Ef} \alpha(\rv),
&
s^y(\rv)
&=
\frac{m I}{\hbar \Ef} \beta(\rv).
\label{eq:spin-galvanic}
\end{align}
This effect is well-known as the inverse of the spin galvanic effect~\cite{Edelstein90, Ganichev02, Ganichev04, Ganichev06}.
In the previous theoretical prediction of the inverse spin galvanic effect in the Rashba system~\cite{Edelstein90}, the constant Rashba coupling was treated non-perturbatively, whereas our calculation is carried out by the perturbation expansion and we take account of the spatial dependence of the Rashba coupling.
Nevertheless these quite distinct calculations are exactly consistent in a condition of the uniform electric current.
(ii) When the limit to the constant spin-orbit coupling is taken first, the spin is not induced identically.
According to the previous work, the inverse spin galvanic effect should occur even in the spatially uniform Rashba system.
It indicates that the inverse spin galvanic effect in the uniform Rashba system is a non-perturbative effect, and thus our result cannot be applicable to the constant Rashba case.

In the above discussion, we omitted the higher-order contribution of electromagnetic field such as the non-commutative contribution of the Yang-Mills magnetic field, $\epsilon^{abc} \Asv^b \times \Asv^c$, in the SU(2)$\times$U(1) Hall effect, $\Iv \times \Bsv^a$.
In fact, this component is related to the spin Hall effect~\cite{Sinova04, Sinitsyn04},
\begin{equation}
s^z(\qv)
=
-\frac{m \kf^2}{\hbar \Ef^2}
(\alpha^2 -\beta^2)
\frac{q_x}{q^2} I(\qv).
\end{equation}
Interestingly the origin of the spin Hall effect is exactly same as the inverse spin galvanic effect, and they are connected with the SU(2)$\times$U(1) Hall effect, $s^a = -(e \tau / m) \div \langle{\Iv \times \Bsv^a}\rangle$.
These effects are classified by whether the non-commutative contribution of the Yang-Mills magnetic field, $\epsilon^{abc} \Asv^b \times \Asv^c$, or the other, $\rot \Asv^a$.

In conclusion, we have analytically derived the general expression of spin polarization arising from electric and magnetic fields in the presence of spin-orbit interaction.
As a result, we obtained the purely electrical spin manipulation, and we have shown that this formula connects different spintronic phenomena which have ever been independently discussed: the inverse spin galvanic effect and the spin Hall effect.
We found also that two different time-dependent Rashba fields yield a large effective magnetic field.
To handle freely the Rashba effect would be a key to the future spintronics.

%%%%%%%%%%

\begin{acknowledgments}
This work was supported by Grant-in-Aid for Scientific Research (S) (Grant No.~24224009) from the Ministry of Education, Culture, Sports, Science and Technology of Japan;
Strategic International Cooperative Program (Joint Research Type) from Japan Science and Technology Agency;
the Funding Program for World-Leading Innovative RD on Science and Technology (FIRST Program).
A.T. is financially supported by the Japan Society for the Promotion of Science for Young Scientists.
\end{acknowledgments}

%%%%%%%%%%

%%%%%%%%%%%%%%%%%%%%

\newpage

\renewcommand{\theequation}{S.\arabic{equation}}
\setcounter{equation}{0}

\begin{center}
%title
{\large \bf
Supplemental Material for
``Theory of Electrical Spin Manipulation in Spin-Orbit Coupling Systems''
}
\\
%authors
{
Akihito Takeuchi$^{1}$ and Naoto Nagaosa$^{1,2}$
}
\\
%affiliations
{\it
$^1$Department of Applied Physics, University of Tokyo, Hongo, Tokyo 113-8656, Japan
\\
$^2$RIKEN Center for Emergent Matter Science, Wako, Saitama 351-0198, Japan
}
\end{center}

%%%%%%%%%%

We will show the details of the analytic calculation of the spin polarization and spin current in the presence of non-Abelian spin-orbit gauge potential and U(1) electromagnetic field.
We carry out the calculation using the Keldysh Green's function based on the quantum many-body theory, and in terms of the Green's function the spin density and spin current are defined as
\begin{equation}
s^a(\rv,t)
=
i e \hbar \tr \big[ \pauli^a \G^<(\rv,t; \rv,t) \big],
\end{equation}
and
\begin{equation}
j_i^a(\rv,t)
=
\frac{i e \hbar}{m} \tr
\bigg\{
\bigg[
\frac{i \hbar}{2} (\grad_{\rv'} -\grad_\rv)_i \pauli^a
+e \A_i(\rv,t) \pauli^a
+e \As_i^a(\rv,t)
\bigg]
\G^<(\rv,t; \rv',t)
\bigg\}_{\rv'=\rv},
\end{equation}
respectively.
In calculation, we consider a disordered regime due to the spin-independent impurity scattering,
\begin{equation}
H_{\rm i}
=
\int{d^d r}\Psi^\dagger(\rv,t) \vi(\rv) \Psi(\rv,t),
\end{equation}
where $\vi$ is the potential of the impurity scattering.
This effect is taken into account as a relaxation time, $\tau$, in the Green's function.
The random impurity averaging is given by ($\ni$ is the impurity concentration and $\ui$ is strength of scattering)
\begin{align}
\overline{\vi(\qv)}
&=
0,
&
\overline{\vi(\qv) \vi(\qv')}
&=
\frac{\ni \ui^2}{L^d} \delta_{\qv,\qv'}.
\end{align}
Thereby we need include the vertex correction shown in Fig.~1(d) for the Ward-Takahashi identity.

%%%%%

\section{SU(2) gauge potential}
We first calculate the electronic spin driven by the effective SU(2) Yang-Mills field due to the spin-orbit interaction.
To confirm rigorously the SU(2) gauge covariance, we consider up to the third-order contribution of the non-Abelian spin-orbit gauge potential.
Therefore, we will obtain the gauge invariant result after the strict calculation.

%%%

\subsection{First order in $\As$}
The diagrammatic representation of the spin density induced by the first order in $\As$ is shown in Fig.~1(a), and this contribution is written down
\begin{align}
s^{(1) a}(\rv,t)
=
&\frac{i 2 e^2 \hbar}{L^d} \sum_{\kv,\qv} \sum_{\w,\W} e^{-i (\qv \cdot \rv -\W t)}
\bigg[
\frac{\hbar}{m} \As_i^a(\qv,\W) k_i 
(f_{\w+\frac{\W}{2}} -f_{\w-\frac{\W}{2}})
\gr_{\kv-\frac{\qv}{2}, \w-\frac{\W}{2}} \ga_{\kv+\frac{\qv}{2}, \w+\frac{\W}{2}} \Pi_\w(\qv,\W)
\notag
\\
&-\Ast^a(\qv,\W)
\big( f_{\w-\frac{\W}{2}} \ga_{\kv-\frac{\qv}{2}, \w-\frac{\W}{2}} \ga_{\kv+\frac{\qv}{2}, \w+\frac{\W}{2}}
-f_{\w+\frac{\W}{2}} \gr_{\kv-\frac{\qv}{2}, \w-\frac{\W}{2}} \gr_{\kv+\frac{\qv}{2}, \w+\frac{\W}{2}} \big)
\notag
\\
&-\Ast^a(\qv,\W)
(f_{\w+\frac{\W}{2}} -f_{\w-\frac{\W}{2}})
\gr_{\kv-\frac{\qv}{2}, \w-\frac{\W}{2}} \ga_{\kv+\frac{\qv}{2}, \w+\frac{\W}{2}} \Pi_\w(\qv,\W)
\bigg],
\end{align}
where $L^d$ is the system size,
$f_\w$ denotes the Fermi distribution function,
$\gr_{\kv,\w}$ ($\ga_{\kv,\w}$) is the impurity-averaged retarded (advanced) Green's function of free electrons defined as ($\Ek = \hbar^2 \kv^2 / 2 m$)
\begin{equation}
\gr_{\kv,\w}
=
\frac{1}{\hbar \w -\Ek +\Ef +\frac{i \hbar}{2 \tau}}
=
(\ga_{\kv,\w})^*,
\end{equation}
and $\Pi_\w(\qv,\W)$ represents contribution of the diffusion ladder
\begin{equation}
\Pi_\w(\qv,\W)
=
\sum_{n=0}^\infty \bigg(
\frac{\ni \ui^2}{L^d} \sum_\kv \gr_{\kv-\frac{\qv}{2}, \w-\frac{\W}{2}} \ga_{\kv+\frac{\qv}{2}, \w+\frac{\W}{2}}
\bigg)^n.
\end{equation}
Considering the slowly varying spin-orbit coupling, $q \ell \ll 1$ and $\W \tau \ll 1$, we carry out the gradient expansion and the leading contribution reads
\begin{align}
s^{(1) a}(\rv,t)
=
&\frac{i 2 e^2 \hbar}{L^d} \sum_{\kv,\qv} \sum_{\w,\W} e^{-i (\qv \cdot \rv -\W t)}
f_\w'
\bigg\{
\frac{i\hbar^3}{m^2} q_j \W \As_i^a(\qv,\W)
\Im \big[ k_i k_j \gr_{\kv,\w} (\ga_{\kv,\w})^2 \big]
\Pi_\w(\qv,\W)
\notag
\\
&-\Ast^a(\qv,\W)
\gr_{\kv,\w} \ga_{\kv,\w}
\bigg[ \frac{i}{\tau} +\W \Pi_\w(\qv,\W) \bigg]
\bigg\}.
\end{align}
After integrating the Green's function with respect to $k$ and $\w$, we finally obtain
\begin{align}
s^{(1) a}(\rv,t)
&=
\cond \sum_\qv \sum_\W
\frac{q_i e^{-i (\qv \cdot \rv -\W t)}}{D \qv^2 +i \W}
\bigg[ \W \As_i^a(\qv,\W) -q_i \Ast^a(\qv,\W) \bigg]
\notag
\\
&=
\frac{\cond}{L^d}
\int{d^d r'} \int{d t'} \sum_\qv \sum_\W
\frac{e^{i \qv \cdot (\rv-\rv') -i \W (t-t')}}{D \qv^2 -i \W}
\grad_{\rv'} \cdot \big[ \del_{t'} \Asv^a(\rv',t') +\grad_{\rv'} \Ast(\rv',t') \big]
\notag
\\
&\equiv
\cond \div \langle{\delt \Asv^a +\grad \Ast^a}\rangle.
\end{align}
The spin current is similarly calculated,
\begin{align}
\jv^{(1) a}(\rv,t)
=
&\frac{i e^2 \hbar^3}{m^2 L^d}
\sum_{\kv, \kv', \qv} \sum_{\w, \W} e^{-i (\qv \cdot \rv -\W t)}
f_\w'
\bigg\{
\notag
\\
&+\frac{i}{6 \tau} (q_i q_j -\delta_{ij} \qv^2)
\As_j^a(\qv,\W)
\gr_{\kv,\w} \ga_{\kv,\w}
+\frac{\hbar^2}{\tau^2}
\W \As_j^a(\qv,\W)
k_i k_j (\gr_{\kv,\w})^2 (\ga_{\kv,\w})^2
\notag
\\
&-\frac{\ni \ui^2}{L^d} \frac{2 \hbar^4}{m^2}
q_k q_l \W \As_j^a(\qv,\W)
\Im \big[ k_i k_k \gr_{\kv,\w} (\ga_{\kv,\w})^2 \big]
\Im \big[ k_j' k_l' \gr_{\kv',\w} (\ga_{\kv',\w})^2 \big]
\Pi_\w(\qv,\W)
\notag
\\
&-i 2 \hbar
q_j \W \Ast^a(\qv,\W)
\Im \big[ k_i k_j \gr_{\kv,\w} (\ga_{\kv,\w})^2 \big]
\Pi_\w(\qv,\W)
\bigg\},
\end{align}
and results in
\begin{equation}
\jv^{(1) a}
=
\chiL \rot (\rot \Asv^a)
-\cond (\delt \Asv^a +\grad \Ast^a)
-D \grad s^{(1) a}.
\end{equation}

%%%

\subsection{Second and third order in $\As$}
Here we consider the higher-order contribution for the SU(2) gauge covariance.
We show the Feynman diagrams of the second- and third-order contributions in Fig.~1(b).
The same manner to the first-order case is applicable to this higher-order case, and each spin polarization is obtained as
\begin{equation}
s^{(2) a}
=
\frac{2 e}{\hbar} \epsilon^{abc}
\Big\langle{
\div
\big[
\cond \Ast^b \Asv^c
+\chiL \Asv^b \times (\rot \Asv^c)
-D \Asv^b s^{(1) c}
\big]
-\Ast^b s^{(1) c}
+\Asv^b \cdot \jv^{(1) c}
}\Big\rangle,
\end{equation}
and
\begin{equation}
s^{(3) a}
=
-\frac{2 e}{\hbar} \epsilon^{abc} \bigg\langle{
\div \bigg[
\frac{e \chiL}{\hbar} \epsilon^{cde} \Asv^b \times (\Asv^d \times \Asv^e)
+D \Asv^b s^{(2) c}
\bigg]
+\Ast^b s^{(2) c}
-\Asv^b \cdot \jv^{(2) c}
}\bigg\rangle,
\end{equation}
respectively.
Correspondingly the spin current is also calculated,
\begin{align}
\jv^{(2) a}
&=
-\frac{2 e}{\hbar} \epsilon^{abc}
\bigg[
\frac{\chiL}{2} \rot (\Asv^b \times \Asv^c)
+\cond \Ast^b \Asv^c
+\chiL \Asv^b \times (\rot \Asv^c)
-D \Asv^b s^{(1) c}
\bigg]
-D \grad s^{(2) a},
\\
\jv^{(3) a}
&=
\frac{2 e}{\hbar} \epsilon^{abc}
\bigg[
\frac{e \chiL}{\hbar} \epsilon^{cde} \Asv^b \times (\Asv^d \times \Asv^e)
+\Asv^b s^{(2) c}
\bigg]
-D \grad s^{(3) a}.
\end{align}
From all the results, the spin and its current densities are represented in a SU(2) gauge invariant form
\begin{align}
s^a
=
&\div \bigg\langle{
\cond \bigg( \delt \Asv^a +\grad \Ast^a +\frac{2 e}{\hbar} \epsilon^{abc} \Ast^b \Asv^c \bigg)}
\notag
\\
&{+\frac{2 e}{\hbar} \epsilon^{abc} \bigg[
\chiL \Asv^b \times \bigg( \rot \Asv^c -\frac{e}{\hbar} \epsilon^{cde} \Asv^d \times \Asv^e \bigg)
-D \Asv^b (s^{(1) c} +s^{(2) c})
\bigg]
}\bigg\rangle
\notag
\\
&-\frac{2 e}{\hbar} \epsilon^{abc}
\big\langle{\Ast^b (s^{(1) c} +s^{(2) c}) -\Asv^b \cdot (\jv^{(1) c} +\jv^{(2) c})}\big\rangle,
\end{align}
and
\begin{align}
\jv^a
=
&\chiL \rot \bigg( \rot \Asv^a -\frac{e}{\hbar} \epsilon^{abc} \Asv^b \times \Asv^c \bigg)
-\cond \bigg( \delt \Asv^a +\grad \Ast^a +\frac{2 e}{\hbar} \epsilon^{abc} \Ast^b \Asv^c \bigg)
\notag
\\
&-\frac{2 e}{\hbar} \epsilon^{abc} \bigg[
\chiL \Asv^b \times \bigg( \rot \Asv^c -\frac{e}{\hbar} \epsilon^{cde} \Asv^d \times \Asv^e \bigg)
-D \Asv^b (s^{(1) c} +s^{(2) c})
\bigg]
-D \grad s^a,
\end{align}
respectively.
This results are rewritten by the effective Yang-Mills field, $\Es$ and $\Bs$,
\begin{align}
s^a
&=
-\cond \div \langle{\Esv^a}\rangle
+\frac{2 e}{\hbar} \epsilon^{abc} \div \langle{\chiL \Asv^b \times \Bsv^c -D \Asv^b s^c}\rangle
-\frac{2 e}{\hbar} \epsilon^{abc} \langle{\As_t^b s^c -\Asv^b \cdot \jv^c}\rangle,
\\
\jv^a
&=
\cond \Esv^a
+\chiL \rot \Bsv^a
-\frac{2 e}{\hbar} \epsilon^{abc}
(\chiL \Asv^b \times \Bsv^c
-D \Asv^b s^c)
-D \grad s^a.
\end{align}
We here introduce the covariant derivative as
\begin{align}
\Delv F^a
&\equiv
\grad F^a -\frac{2e}{\hbar} \epsilon^{abc} \Asv^b F^c,
\\
\Delt F^a
&\equiv
\delt F^a +\frac{2e}{\hbar} \epsilon^{abc} \Ast^b F^c,
\end{align}
where $F^a$ is an arbitrary function in spin space.
The spin current is simplified using this covariant derivative,
\begin{equation}
\jv^a
=
\cond \Esv^a +\chiL \Delv \times \Bsv^a -D \Delv s^a,
\end{equation}
and the spin polarization is given by the covariant conservation law of spin,
\begin{equation}
\Delt s^a = -\Delv \cdot \jv^a.
\end{equation}

%%%%%

\section{SU(2) and U(1) gauge potential}
Next, we calculate the spin arising from a combination between the effective SU(2) Yang-Mills field and the usual U(1) Maxwell electromagnetic field.
The diagrams of this contribution is shown in Fig.~1(c).
Although the calculation becomes more and more complicated, spin and spin current densities are straightforwardly derived,
\begin{align}
s^a
=
&\frac{e \tau \cond}{m} \div \bigg\langle
(\rot \Av) \times \Big[ \delt \Asv^a +\grad \Ast^a
+D \grad \big\langle{\div (\delt \Asv^a +\grad \Ast^a)}\big\rangle \Big]
\notag
\\
&+(\rot \Asv^a) \times \Big[ \delt \Av
+\grad \At +D \grad \big\langle{\div (\delt \Av +\grad \At)}\big\rangle \Big]
\notag
\\
&-(\delt \Av +\grad \At) \big\langle{\div (\delt \Asv^a +\grad \Ast^a)}\big\rangle
-(\delt \Asv^a +\grad \Ast^a) \big\langle{\div (\delt \Av +\grad \At)}\big\rangle
\notag
\\
&+\frac{1}{D} \At \Ast^a
+\grad \Big[ \At \big\langle{\div (\delt \Asv^a +\grad \Ast^a)}\big\rangle \Big]
+\grad \Big[ \Ast^a \big\langle{\div (\delt \Av +\grad \At)}\big\rangle \Big]
\bigg\rangle,
\\
\jv^a
=
&-\frac{e \tau \cond}{m} \bigg\{
(\rot \Av) \times \Big[ \delt \Asv^a +\grad \Ast^a
+D \grad \big\langle{\div (\delt \Asv^a +\grad \Ast^a)}\big\rangle \Big]
\notag
\\
&+(\rot \Asv^a) \times \Big[ \delt \Av
+\grad \At +D \grad \big\langle{\div (\delt \Av +\grad \At)}\big\rangle \Big]
\notag
\\
&-(\delt \Av +\grad \At) \big\langle{\div (\delt \Asv^a +\grad \Ast^a)}\big\rangle
-(\delt \Asv^a +\grad \Ast^a) \big\langle{\div (\delt \Av +\grad \At)}\big\rangle
\notag
\\
&+\frac{1}{D} \At \Ast^a
+\grad \Big[ \At \big\langle{\div (\delt \Asv^a +\grad \Ast^a)}\big\rangle \Big]
+\grad \Big[ \Ast^a \big\langle{\div (\delt \Av +\grad \At)}\big\rangle \Big]
\bigg\}
-D \grad s^a.
\end{align}
This result expressed by the SU(2) and U(1) gauge potentials has a lot of contribution and it is obscure to explain phenomena.
To replace each gauge potential with the effective Yang-Mills field and the electromagnetic field, the equation is simplified as
\begin{align}
s^a
&=
-\frac{e \tau}{m}
\div
\Big\langle{
\Bv \times \Isv^a
+\Bsv^a \times \Iv
-S^a \Ev
-\rho \Esv^a
-\grad \big( e^2 \nu \At \Ast^a +\At S^a +\rho \Ast^a \big)
}\Big\rangle,
\\
\jv^a
&=
\frac{e \tau}{m}
\Big[
\Bv \times \Isv^a
+\Bsv^a \times \Iv
-S^a \Ev
-\rho \Esv^a
-\grad
\big( e^2 \nu \At \Ast^a +\At S^a +\rho \Ast^a \big)
\Big]
-D \grad s^a.
\end{align}
In this calculation, we considered the linear response of the non-Abelian spin-orbit gauge potential.
The effective Yang-Mills field does not contain non-commutative parts proportional to $\epsilon^{abc} \Ast^b \Asv^c$ and $\epsilon^{abc} \Asv^b \times \Asv^c$, and therefore the field turns to $\Esv^a = -\delt \Asv^a -\grad \Ast^a$ and $\Bsv^a = \rot \Asv^a$.
However, the non-commutative contribution should exist at the viewpoint of the SU(2) gauge covariance.
Since the result depends on the first-order spin-orbit coupling, spin polarization and spin current given by Eqs.~(S.26) and (S.27) are exactly conserved,
\begin{equation}
\delt s^a +\div \jv^a
=
0.
\end{equation}
Considering the non-commutative parts in the Yang-Mills field, the derivative of this identity is surely replaced by the covariant derivative.

\end{document}